\documentclass[preprint,review,12pt,sort&compress]{elsarticle}
\usepackage[margin=1.18 in]{geometry}
\usepackage{multirow}

\usepackage{amsmath}
\usepackage{amssymb}
\usepackage{gensymb}
\usepackage{bm}
\usepackage{enumitem}
\usepackage{lipsum}
\usepackage{makecell}
\usepackage{float}
\DeclareMathOperator*{\argmin}{arg\,min}

\usepackage{xcolor}

\journal{Composite Science and Technology}

\begin{document}

\begin{frontmatter}

%% Title, authors and addresses

%% use the tnoteref command within \title for footnotes;
%% use the tnotetext command for theassociated footnote;
%% use the fnref command within \author or \address for footnotes;
%% use the fntext command for theassociated footnote;
%% use the corref command within \author for corresponding author footnotes;
%% use the cortext command for theassociated footnote;
%% use the ead command for the email address,
%% and the form \ead[url] for the home page:
%% \title{Title\tnoteref{label1}}
%% \tnotetext[label1]{}
%% \author{Name\corref{cor1}\fnref{label2}}
%% \ead[url]{home page}
%% \fntext[label2]{}
% \cortext[cor1]{Corresponding Author, \ead{salviato@aa.washington.edu}}
\cortext[cor1]{Corresponding Author}
\fntext[econt]{Equally contributed first author.}
%% \address{Address\fnref{label3}}
%% \fntext[label3]{}
% \jk{Point of discussion: \\
% 1. Shall we replace `non-conventional' with `generalized' or `double-double'?\\
% 2. Parameters in Table 1 need to be described. For example, we need to emphasize that detailed calculation has been made, such as head travel and turn time, for the calculation of `t'. \\ 
% 3. Please check if the image in Fig. 1A is free to use. Also, if possible, I would rather find an image that places a tape on a flat bed, not a curved one.\\
% 4. Figures 2(a) and (b) are basically the same, aren't they? If so, please merge them into one.\\
% 5. All figures and tables need to be mentioned and explained in the text, and they should be introduced in order. Particularly, you have created beautiful figures 4 and 5, but you never mention them in the manuscript!!!\\
% 6. Overall, the description on our sophisticated optimization method has not been emphasized enough. The random walk like optimization steps shown in Figure 4 are so amazing, but we have not talked about it...}

% \ky{Dr. Yang, thank you for your comments:\\
% 2. Sean added the explanation, thanks!\\
% 3. and 4. What do yo think, Sean?\\
% 5. I modified the manuscript and currently they are mentioned in order.\\
% 6. I added more emphasis about Fig 4 in the conclusion.\\
% I also added Dr. Tsai's book to bib file.}

\title{Ply-drop design of non-conventional composites using Bayesian optimization}

%% use optional labels to link authors explicitly to addresses:
%% \author[label1,label2]{}
%% \address[label1]{}
%% \address[label2]{}

\author[address]{Koshiro Yamaguchi}
\author[address]{Sean E. Phenisee\fnref{econt}}
\author[address]{Zhisong Chen}
\author[address]{Marco Salviato\corref{cor1}}
\ead{salviato@aa.washington.edu}
\author[address]{Jinkyu Yang}

\address[address]{William E. Boeing Department of Aeronautics and Astronautics, University of Washington, Seattle, Washington 98195-2400, USA}

\begin{abstract}
\linespread{1}\selectfont
%% Text of abstract
Automated Fiber Placement (AFP) technology provides a great ability to efficiently produce large carbon fiber reinforced composite structures with complex surfaces. AFP has a wide range of tow placement angles, and the users can design layup angles so that they can tailor the performance of the structure. However, despite the design freedom, the industry generally adopts a layering of $0 \degree$, $90 \degree$, and $\pm 45 \degree$ ply-drop angles. Here, we demonstrate the optimization of ply-drop angles of non-conventional composites. Specifically, we use classical laminate theory and Bayesian optimization to achieve better layup angles in terms of stiffness, Tsai-Wu failure criteria, and manufacturing time. Our approach shows its effectiveness in designing carbon fiber composite structures using unconventional angles in terms of both mechanical properties and production efficiency. Our method has the potential to be used for more complex scenarios, such as the production of curved surfaces and the utilization of finite element analysis.
\end{abstract}

\begin{keyword}
Automated Fiber Placement \sep Laminate theory\sep Failure criterion \sep Manufacturing properties \sep Bayesian optimization
%% keywords here, in the form: keyword \sep keyword

%% PACS codes here, in the form: \PACS code \sep code

%% MSC codes here, in the form: \MSC code \sep code
%% or \MSC[2008] code \sep code (2000 is the default)

\end{keyword}

\end{frontmatter}
%% \linenumbers

%% main text

\section{Introduction}
\label{intro}
% Koshiro wrote:
The prominent advantage of fiber-reinforced composites is their outstanding mechanical properties in terms of specific stiffness and strength which have fostered the use of these materials not only in the field of military aircraft but also in commercial aircraft and automotive. Another distinctive feature of fiber composites compared to metal alloys is the anisotropic mechanical behavior that enables a more efficient optimization of the structural performance. Fiber composites have great flexibility of design, for example, in terms of fiber-matrix combinations, in-plane ply arrangements, and out-of-plane layup sequences. However, while this represents one of the greatest benefits of composites, it also brings complexity to determine the optimal design of composite structures. To overcome such complexity, the industry normally employs a limited set of layup angles such as $0 \degree$, $90 \degree$, and $\pm 45 \degree$.
% careful on a strong claim on QI usage in industry
This approach can simplify the design and manufacturing process of composite materials. However, it also hinders the superb mechanical properties of composites with tailored design compared to the metal counterparts (e.g. employing unconventional layup angles for composite wings \cite{kim2017}). 

\indent Automatic Fiber Placement (AFP) machines are extending the limits of the manufacturing of composite materials to fully exploit their design freedom \cite{Grimshaw2001, DirKev12, FrkRam17}. They place a narrow tape of pre-pregs in a strategic way to achieve tailored structural profiles by areas. 
% Sean wrote:
% AFP studies on defect, process modeling and optimization ------
% \jk{(JK: It is a great job done here to describe the defect studies in the setting of AFP-manufactured composites. However, our study does not address composite defects. So it can be misleading by having this literature survey here. I recommend removing this highlighted part. We can save this paragraph for our future studies.) [delete from here] [delete up to here.]}
% \sean{Defect studies removed}
Taking advantage of the advancement in manufacturability provided by AFP technology, optimization studies have been conducted to produce optimal composite laminates that can outperform conventional composite structures. Nik et al. \cite{NikLes14} performed multi-objective optimization on curvilinear fiber paths to maximize both buckling load and stiffness of the laminate with embedded defects under a uniaxial compression; Vijayachandran et al. \cite{VijWaa20} optimized fiber paths to maximize the critical buckling load of a flat square plate under the biaxial in-plane compression, accounting for the effect of gaps and overlaps induced by AFP manufacturing. Tsai et al \cite{Tsai2019} recently proposed so-called double-double laminates (i.e., laminates with $[\pm \phi / \pm \psi]$ layup sequences) to facilitate the production efficiency of composites without sacrificing their strength in certain loading conditions. 

% ---------------------------------------------------------------

% Koshiro wrote:
Since AFP-based production is highly automated, we can also account for the operating time as the production cost. Considering the slow production rate of commercial aircraft even with the use of AFP machines, optimization on production time is also of importance. However, the current AFP software is mainly for the planning of operation paths of AFP without thoroughly considering multidimensional values such as structural performance and operation time. Here, we tackle this complex problem of AFP-based composite design and manufacturing. This task is challenging due to the multi-objective nature of the design optimization. That is, as we consider \textit{n}-number of variations for the total of \textit{m}-parameters, the total number of simulations becomes $n^m$. This is a so-called \textit{curse of dimensionality}. For the design of composite structures such as aircraft wing, this number becomes astronomical, and the conventional approach of optimization would not be able to identify the optimal solution.

Recently, several researchers have tackled this problem for materials design with the aid of machine learning and optimization technology \cite{Bessa2017}. They demonstrated that the machine-learning-based approach for material/structural design driven by data from numerous simulations is a promising technique. Herein, we further leverage this technology into the manufacturing stage. Our data-driven optimization approach refers to a recent advance to design slender structures and metamaterials \cite{Bessa2018, Bessa2019}. This framework combines the design of experiments, efficient analyses to evaluate the output, and optimization to achieve ideal designs. Specifically, we use Bayesian optimization methods, not only to account for structural performance and manufacturing efficiency, but also to make our framework sufficiently scalable for other factors, e.g., probabilistic manufacturing imperfections and high-cost structural analysis computations. As proven by the success of recent data-driven attempts, we anticipate that this approach will achieve better performance in the proposed application of non-conventional composite manufacturing using AFP machines.

Although manufacturing optimization is a highly complex problem, extensive researches have been conducted to overcome the challenges related to the design of composite structures. For example, thickness optimization of blended composite structures \cite{FarzanNasab2018} and machine learning approaches for the AFP manufacturing \cite{Bruning2017} are notable ones. However, these studies are in lack of the design scheme of layup angles. The former tries to optimize the number of layups with the pre-defined set of layup angles $(0\degree,90\degree,\pm 45\degree)$ and the latter focuses on the process parameters (e.g. layup velocity, heater temperature, and compaction pressure) and manufacturing defects. The optimization study carried out by Nik et al. \cite{NikLes14} focuses on improving structural performance by taking advantage of variable stiffness. However, this study does not consider a manufacturing time in the design variable. In this study, we demonstrate the application of Bayesian optimization and classical laminate theory to design unconventional composite layups to achieve better mechanical and production properties in terms of stiffness, Tsai-Wu failure criteria, and manufacturing time. Our approach shows its effectiveness in designing carbon fiber composite that fulfills both mechanical properties and production efficiency. This approach has great potential for designing composite materials with more complex geometries where each function evaluation during optimization requires computationally heavy methods such as finite element analysis. 
% \jk{(Is `observation' a proper word here?) } \ky{the word "observation" is often used in the literature of BO. However, "function evaluation" might be good for the readers. So I rephrased it.}

\section{Productivity and structural module development}
%Under the development of the design guide tool, 
We developed a comprehensive design guide module that evaluates both the AFP operation efficiency and the structural performance of the laminate. Tape-layering simulation provides measures of production cost such as a total manufacturing time and the area of cut-out material waste. In addition, the structural analysis component of the module examines the stiffness and strength of the laminated product. Theses combined evaluation results serve as an input for the optimization process to produce an optimal candidate for the composite design.

\subsection{Tape-layering simulation}
\begin{figure} [H]
\center
\includegraphics[scale=0.55]{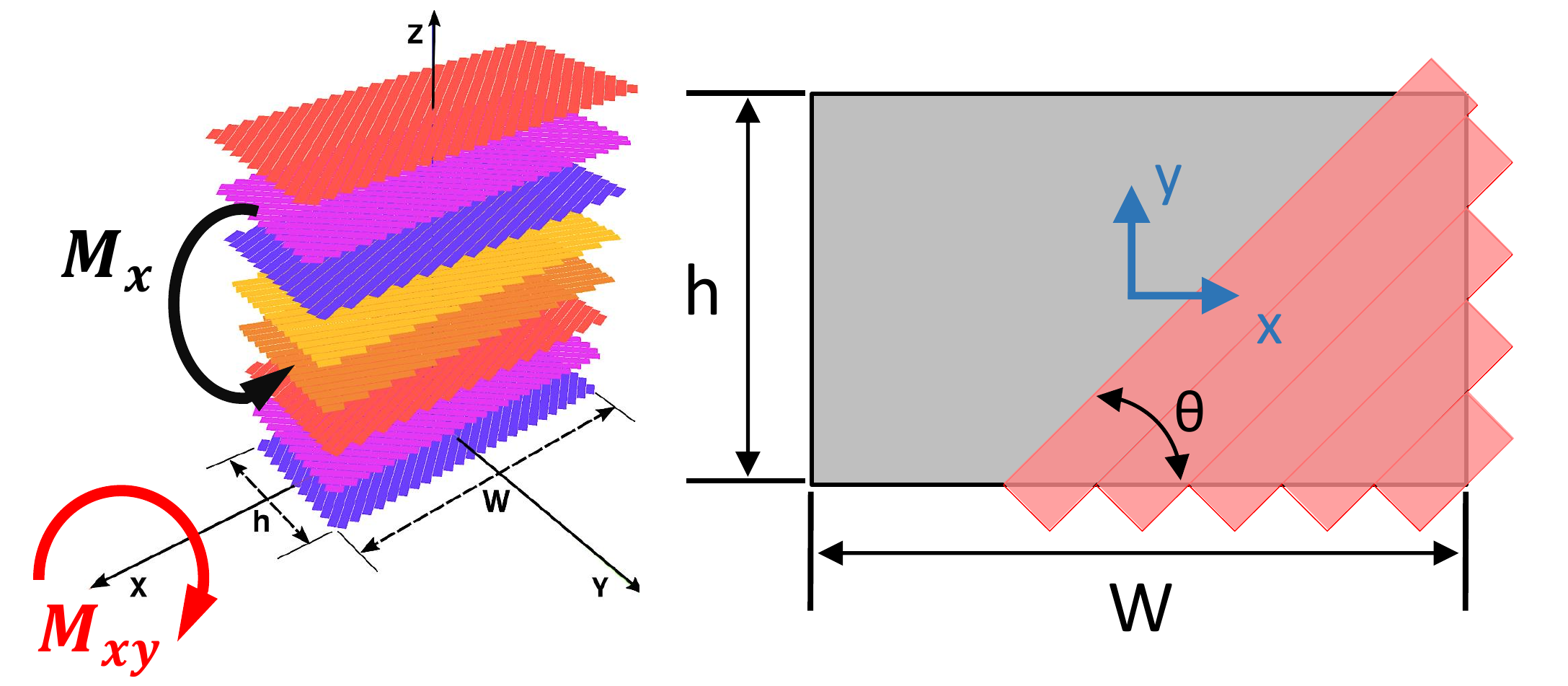}
\caption{Modeling the tape-layering paths for a rectangular plate geometry}
\label{manufacturing_model}
\end{figure}

For the simplicity of the analysis, we chose a flat rectangular plate as a surface profile, and fiber paths remain straight on the surface. We did not include potential manufacturing defects such as overlaps and gaps since the focus of this study is not the degradation in mechanical properties due to defects. Inputs for the tape-layering simulation are dimensions of a rectangular plate, the angle of the ply, tow width, tow drop number, the speed of the machine head, and its rotational speed while it moves to the next set of tows. See Table 1 for the input parameters used in this study. Given that these parameters are defined, the simulation produces manufacturing parameters such as operation time and the area of cut-out materials which represents the measure of wasted materials for the specified ply. A visualization of the lay-up simulation is shown in Fig. \ref{manufacturing_model}. A simulation is performed for each ply of the laminate. The summation of the results for each ply provides total production time and material waste for the laminate, which is fed up in the optimization process.

\subsection{Evaluation of structural performance}
For the structural analysis, we focused on two parameters: stiffness and strength of the laminate. Classical Laminate Theory was used to compute the stiffness of the laminate, and Tsai-Wu criterion with the invariant concept was used to compute the strength measure of the laminate that approximates the initial ply failure.
\subsubsection{Classical Laminate Theory}
We assume that the laminate is thin compared to its lateral dimension and out-of-plane components of stress tensor are negligible relative to the in-plane components. Furthermore, Kirchhoff hypothesis holds for the plate. Therefore, there is no significant out-of-plane shear deformation, and the straight lines perpendicular to the mid-surface stays straight and perpendicular after the plate is deformed under in-plane loading. 

Following the Classical Laminate Theory \cite{KolSpr03}, in-plane forces and moments of a laminate relate to the reference plane strain components and curvatures of the laminate through $[A]$, $[B]$ and $[D]$ matrices, which represent the stiffness of the laminate. 
In this study, we focused on the $D_{11}$ and $D_{66}$ of $[D]$ matrix. They relate to the stiffness of the laminate in bending moment and in-plane shear due to  torsion, respectively (directions shown in Fig. \ref{manufacturing_model}). In this study, these two loading types are the main interests in evaluating the structural performance of the laminate considering aerospace applications. 

\subsubsection{Failure criterion}
We use tensor polynomial based failure criterion: $G_{ij}\epsilon_{i}\epsilon_{j} + G_i \epsilon_i= 1$, widely known as Tsai-Wu criterion in order to predict the first-order estimation on the initial failure of a laminate. $G_{ij}$ and $G_i$ represent strength parameters in strain space under the plane stress assumption. Detailed steps for computing the parameters are referred to Tsai and Melo. \cite{TsaMel14}. The choice of this criterion is motivated by its effectiveness and accuracy in capturing the failure condition of smooth composite laminates. However, for more complex structural configurations and loading conditions, the emergence of large Fracture Process Zones (FPZs) can lead to significant size effects. This is typically the case of structures featuring open and filled holes, notches and other stress raisers. In such a case, a quasibrittle fracture mechanics framework should be preferred over a stress-based failure criterion \cite{bazant, salviato1, salviato2, salviato3, Okabe}. We construct failure envelops using strain space on account of the highly anisotropic property of carbon fiber reinforced composites since the shape of envelopes in strain space is invariant. 
% \jk{(The highlighted part here can be deleted.) However, for more complex structural configurations and loading conditions, the emergence of large Fracture Process Zones (FPZs) can lead to significant size effects. This is typically the case of structures featuring open and filled holes, notches and other stress raisers. In such a case, a quasibrittle fracture mechanics framework should be preferred over a stress-based failure criterion \cite{bazant, salviato1, salviato2, salviato3, Okabe}.} 
% \ms{If possible, I would like to keep this part. It is important to distinguish between the two failure condition and highlight the appropriate failure criteria.} 

\begin{figure} [H]
\center
\includegraphics[scale=0.65]{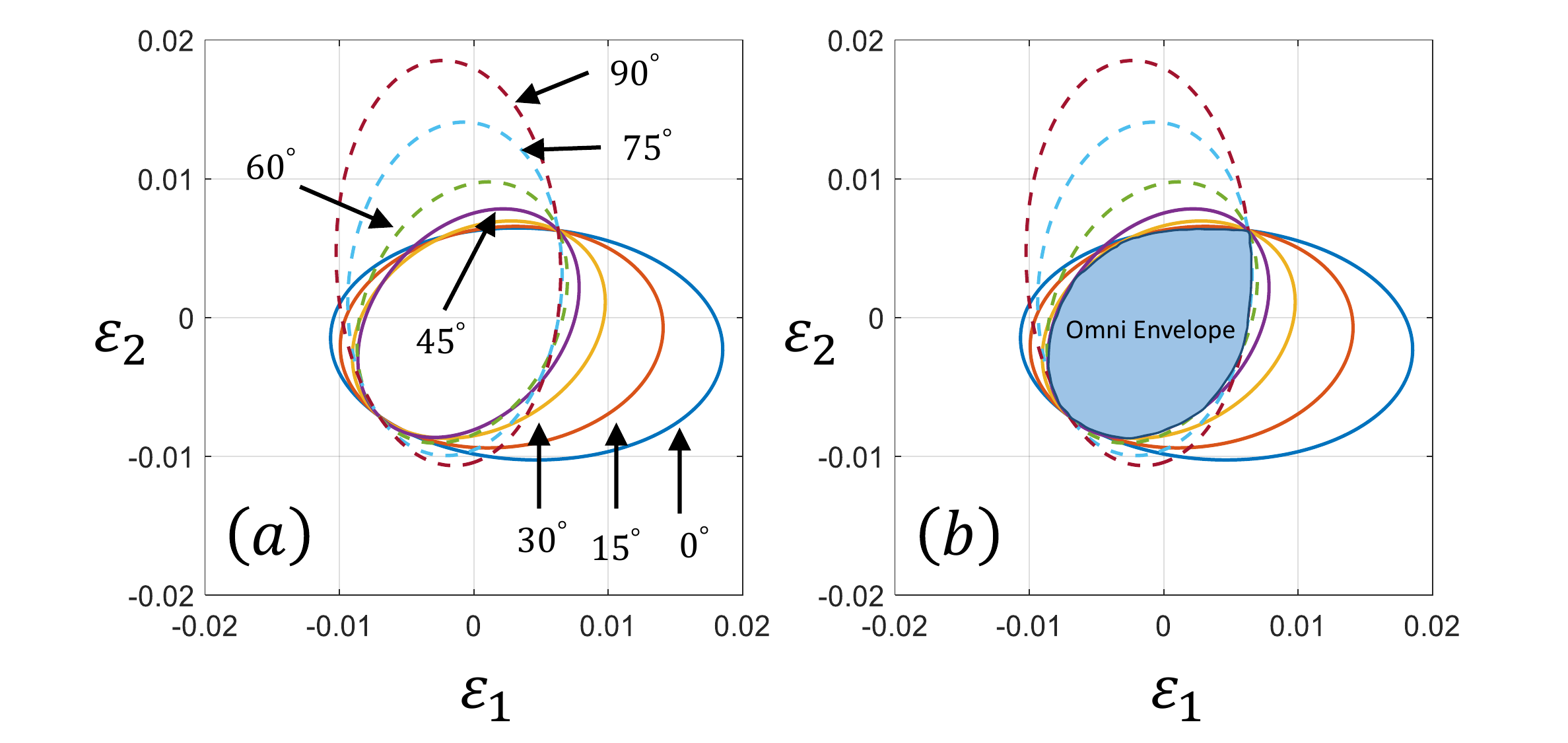}
\caption{Tsai-Wu failure envelopes in strain space for IM7-977 are shown in $(a)$ for multiple angles. Omni strain envelope is highlighted for the laminate in $(b)$.}
\label{failure_envelope}
\end{figure}

Tsai-Wu strain envelops of multiple angles for IM7-977 material system are shown in Fig. \ref{failure_envelope} (a). Hypothetically, we can determine the inner envelope, so-called omni-envelope, by taking the intersection of all the envelops that sweep the ply angle from $0^{\circ}$ to $90^{\circ}$. By construction, the omni-envelop is a layup sequence invariant and a superior tool to predict the initial ply failure. Fig. \ref{failure_envelope} (b) shows the omni-envelope of IM7-977. To closely adopt the idea of omni-envelop, we compute the Tsai-Wu value for each ply and find the maximum value in the laminate that represents the layer likely to reach initial failure. In the optimization process, we use the maximum Tsai-Wu value of the laminate as a strength indicator of the laminate.   

% Tsai-Wu - strain space
% introduce concept of invariant envelope

% \section{Mathematical framework (need to be renamed?)}

\section{Optimization Algorithm}
We adopt the Bayesian optimization approach to tailor the ply-drop angles of non-conventional composites. Generally, there are two types of Bayesian methods that can be applied to the design of advanced materials such as composites or mechanical metamaterials. The one is Bayesian machine learning \cite{Ghahramani2015}, and the other is Bayesian optimization \cite{Gonzalez2016}. While the former technique helps us obtain a complete map of the solution space, the latter efficiently searches one optimum solution for a particular purpose through the adaptive sampling of the solution space. Therefore, Bayesian optimization suits our goal in the best way and is a derivative-free and effective method for global optimization of multimodal objective functions with a high cost of the evaluation \cite{Lam2018}. Bayesian optimization utilizes all the past information acquired to build a surrogate model. The algorithm uses this surrogate model to choose the next design for evaluation. Since it is a derivative-free method, we can potentially apply the same optimization framework for optimizations on composite structures with complex geometries via evaluating them with finite element analysis. 

The Bayesian optimization algorithm works on the following optimization problem:

\begin{equation}
    \bm{x}^* = \argmin_{\bm{x}\in \chi}f(\bm{x})
\end{equation}

\noindent Here, $\argmin f(\bm{x})$ is the value of $\bm{x}$ for which $f(\bm{x})$ attains its minimum. Objective function $f$ is a function that takes a $d$-dimensional input vector $\bm{x}$ from a design space $\chi \subset \mathbb{R}^d$ and usually is expensive to evaluate. We compute a minimizer $\bm{x}^*$ of $f$. Here, the minimizer is not always unique. The first step of Bayesian optimization is to sample an initial training set $\mathcal{D}_1 = \{(\bm{x}_1,f(\bm{x_1}))\}$ that contains a input vector $\bm{x}_1$ and a value of objective function $f(\bm{x}_1)$. For each iteration $n$, we construct a statistical model, a Gaussian process \cite{Rasmussen2006}, from the training set $\mathcal{D}_n$. Then, the posterior mean $\Bar{\mu}(\bm{x};\mathcal{D}_n,f)$ of the Gaussian process related with $f$ at $x$ conditioned on  $\mathcal{D}_n$ works as a surrogate for the objective function $f$. Likewise, the posterior variance $\Bar{\sigma}^2(\bm{x};\mathcal{D}_n,f)$ is a measure of the uncertainty of the surrogate of the objective function $f$. Based on the statistical surrogate model, a utility function $\mathit{U}_n(\bm{x};\mathcal{D}_n)$ quantifies the benefits of evaluation for a new design $\bm{x}$ according to the surrogate model built with $\mathcal{D}_n$. The next design, $x_{n+1}$, is obtained by solving an adjacent optimization problem that maximize the utility function. Here, the Expected Improvement (EI), the Probability of Improvement (PI), the Upper Confidence Bound (UCB) and the Lower Confidence Bound (LCB) are notable utility functions \cite{Snoek2012}. We adopt the LCB for our optimization program. This process repeats until a stopping criterion is satisfied (e.g. number of allowed iteration $N$). This is the overview of a single-objective Bayesian optimization. 
However, in many real-world applications such as this problem, they can be formulated as multi-objective optimization problems. We aim to optimize the system for multiple criteria. In such a case, it is unlikely to be able to optimize all of the criteria simultaneously because they usually conflict with each other. Still, it is possible to find a set of optimal points known as the Pareto set \cite{Murata1996}. This set contains a collection of solutions where no objective can be improved without compromising others. In this paper, we utilize a simple scalarization technique with normalization of each objective to find one solution in a Pareto set because even within this scheme, we can show a better solution of lay-up angles compared to quasi-isotropic layering. To implement this Bayesian optimization framework, we develop Python codes for the evaluation of mechanical properties, manufacturing time, and optimization process with GPyOpt \cite{gpyopt2016}.  

\begin{figure} [H]
\center
\includegraphics[scale=0.85]{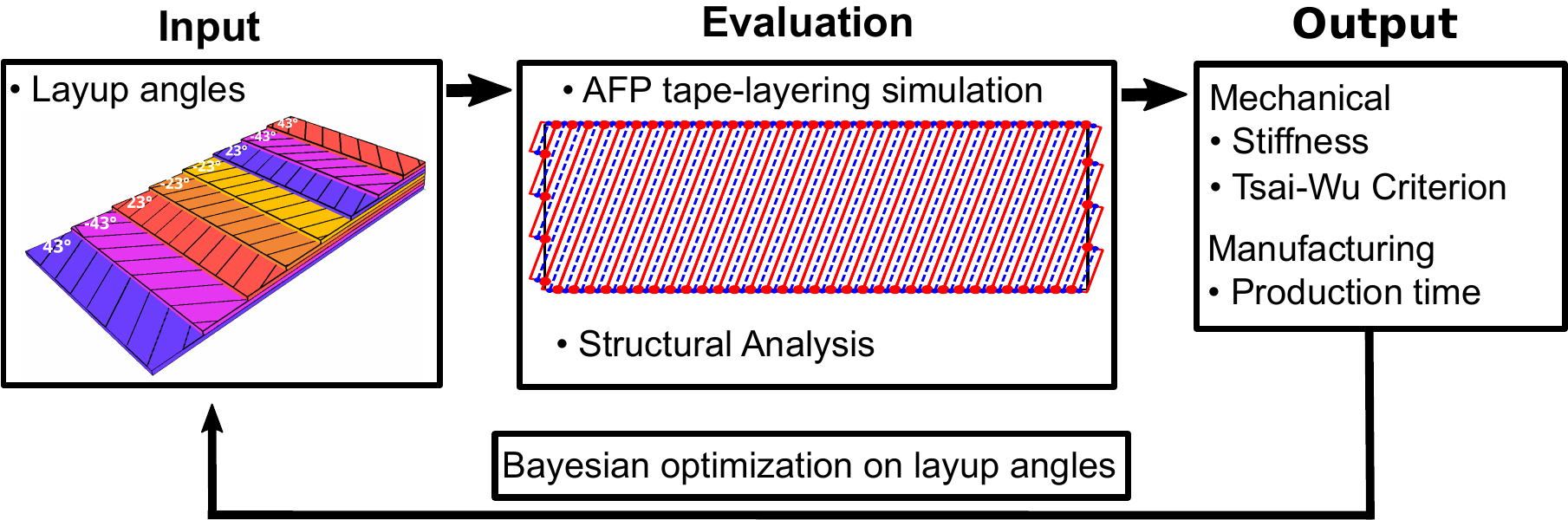}
\caption{Design of the optimization process}
\label{figure_schematic_new}
\end{figure}

Fig. \ref{figure_schematic_new} summarizes the schematic of our optimization framework. Here, the target variables we want to achieve optimization of are layup angles for a laminate. Based on those layup angles, we perform the tape-layering simulation and the evaluation of structural performance through the classical laminate theory and Tsai-Wu critrion. As the outputs of the evaluation, we obtain four values: $D_{11}$ and $D_{66}$ as stiffness, $TW$ as the maximum Tsai-Wu value for each set of layup angles, and $t$ as production time. To combine those metric into one objective function, we apply scalarization with weights and normalization. We prepared twelve sets of weight values with three different loading conditions to observe how our framework behaves with various scenarios. Table \ref{tab:Table_weights} shows all weight values with loading conditions. Here, Condition 1 is $M_x=10[\text{Nmm}]$ and $M_{xy}=1[\text{Nmm}]$. Condition 2 is $M_x=1[\text{Nmm}]$ and $M_{xy}=10[\text{Nmm}]$. Condition 3 is $M_x=10[\text{Nmm}]$ and $M_{xy}=10[\text{Nmm}]$, where $M_x$ and $M_{xy}$ are bending moments caused by axial stresses in $x$-direction and shear stresses in $xy$-direction, respectively. Other loading directions are all zero.

\begin{figure} [H]
\center
\includegraphics[scale=0.9]{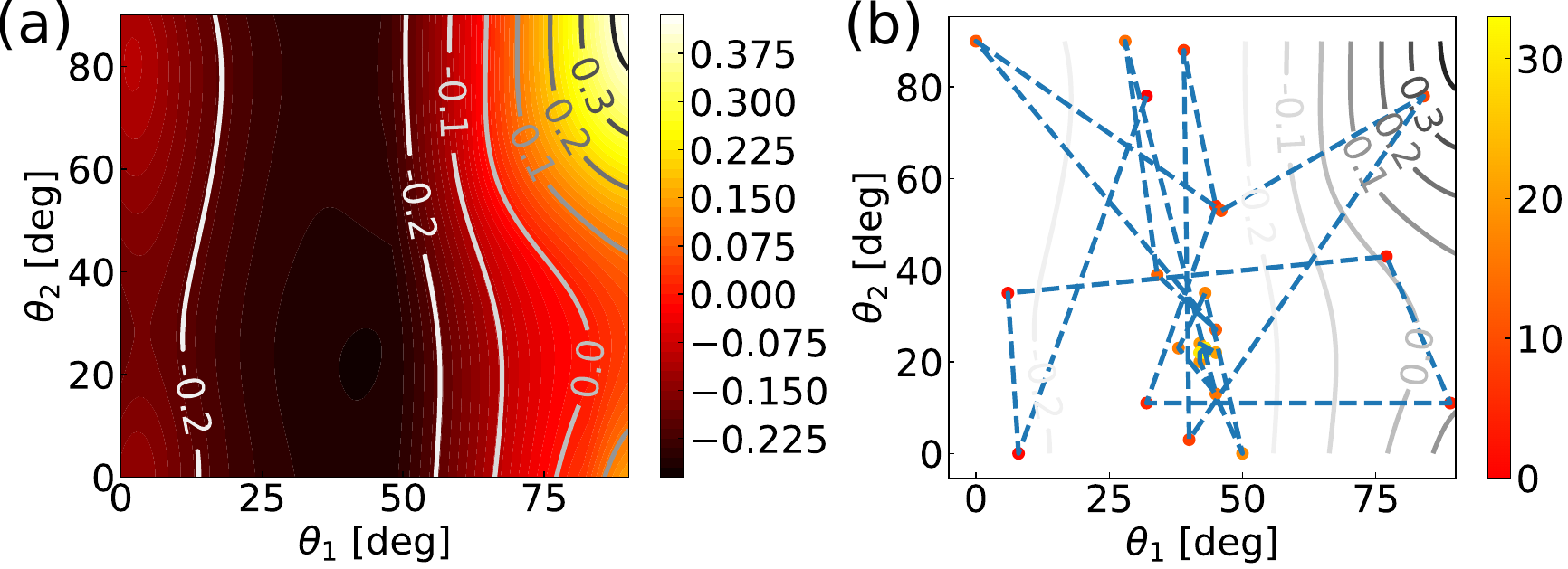}
\caption{Contour plot of cost function and process of Bayesian optimization. (two-variable optimization of an 8-ply layup, Case 6) Left panel (a) is a contour plot generated computationally. Right panel (b) shows the process of the optimization. The contour plot on the right figure is the same as the left one. Dots are the data points acquired through the optimization. Red dots are obtained in the earlier stage of the optimization, whereas the yellow ones are calculated in the later stage.}
\label{figure_landscape}
\end{figure}

First we consider 8-ply balanced and symmetric laminate. It can be expressed as $\bm{\theta}=[\pm \theta_1, \pm \theta_2]_s$. The assumption of the angle of the layup angle for automated fiber placement machines is between $-90 [\text{deg}]$ and $90 [\text{deg}]$. However, due to the symmetry of the laminate, we limit $\theta_1$ and $\theta_2$ between $0 [\text{deg}]$ and $90 [\text{deg}]$. To normalize each variables, we use the maximum values for each one of them. $D_{11max}$ is the value of a laminate $[0]_8$, $D_{66max}$ and $TW_{max}$ are the value of a laminate of $[45,-45]_{2s}$, $t_{max}$ is the value of a laminate of $[67,-67]_{2s}$. This optimization problem can be formulated as in equation \eqref{eq:1}.
\begin{equation}
    \begin{split}
        & \text{minimize}\ f(\theta_1,\theta_2)=-w_1 \frac{D_{11}}{D_{11max}} - w_2 \frac{D_{66}}{D_{66max}} + w_3\frac{TW}{TW_{max}} + w_4\frac{t}{t_{max}}\\
        \text{where}\\
        &\{(\theta_1,\theta_2) \in \mathbb{Z} | 0 \leq \theta_1,\theta_2 \leq 90\}.\\
    \end{split}
    \label{eq:1}
\end{equation}
% \jk{(TW and t have never been introduced earlier. So you need to clearly define what the represent.)} \ky{updated}
Here, $w_1$ to $w_4$ are the weights assigned to each optimization case as described in Table \ref{tab:Table_weights}. Since it is a two-dimensional problem, it is easy to visualize the process and we can speculate how the algorithm works. For example, the contour plot and the visualization of the optimization process for Case 6 are shown in Fig.\ref{figure_landscape}.
Here we find that the cost function f varies as we use a different combination of ply-drop angles: $\theta_1$ and $\theta_2$. The dark area in Fig. \ref{figure_landscape}(a) is the region that exhibits the minimal $f$, implying the optimized configuration of the eight-ply composite under case 6 scenario as described in Table \ref{tab:Table_weights}. The process of finding this solution via Bayesian optimization is shown in Fig. 4(b). We started with the initial guess of $(\theta_1, \theta_2) = (31, 78)$, marked in red in Fig. \ref{figure_landscape}(b). Through the iterations, various combinations of the two variables are attempted, but eventually, the solution approaches $(\theta_1, \theta_2) = (43, 23)$ after 31 iterations. This optimal solution is in agreement with the result in the contour map in Fig. \ref{figure_landscape}(a). For different simulation cases as listed in Table \ref{tab:Table_weights}. 
The result of the optimization is shown in Table \ref{tab:Table_result8ply_case123} to Table \ref{tab:Table_result8ply_case789}. Overall, the simulation result shows how Bayesian optimization works well with
multimodal functions as a derivative-free method. The discussion on each case will be in Section 4.

To assess the efficacy of the proposed optimization scheme for a large number of plies (i.e., more than two-variable optimization case), we also apply this algorithm to a 32-ply layup. In this case, it can be expressed as $\bm{\theta} = [\pm \theta_1,\pm \theta_2, \pm \theta_3,\pm \theta_4, \pm \theta_5, \pm \theta_6, \pm \theta_7, \pm \theta_8]_s$. This problem is formulated as:
\begin{equation}
    \begin{split}
        & \text{minimize}\ f(\bm{\theta})=-w_1 \frac{D_{11}}{D_{11max}} - w_2 \frac{D_{66}}{D_{66max}} + w_3\frac{TW}{TW_{max}} + w_4\frac{t}{t_{max}}\\
        \text{where}\\
        &\{(\theta_1,\theta_2,\theta_3,\theta_4,\theta_5,\theta_6,\theta_7,\theta_8) \in \mathbb{Z} | 0 \leq \theta_1,\theta_2,\theta_3,\theta_4,\theta_5,\theta_6,\theta_7,\theta_8 \leq 90\}.\\
    \end{split}
    \label{eq:2}
\end{equation}

% (Describe MOO(maybe go to SI?))
% Though we show the effectiveness of the Bayesian optimization scheme to improve the both mechanical performance and productivity of composite materials with a simple scalarization technique with normalization and adding weights. However, in order to fully explore the design space of the composites, we apply ParEGO algorithm to our problem.

\section{Results and discussion}
\begin{figure} [H]
\center
\includegraphics[scale=0.6]{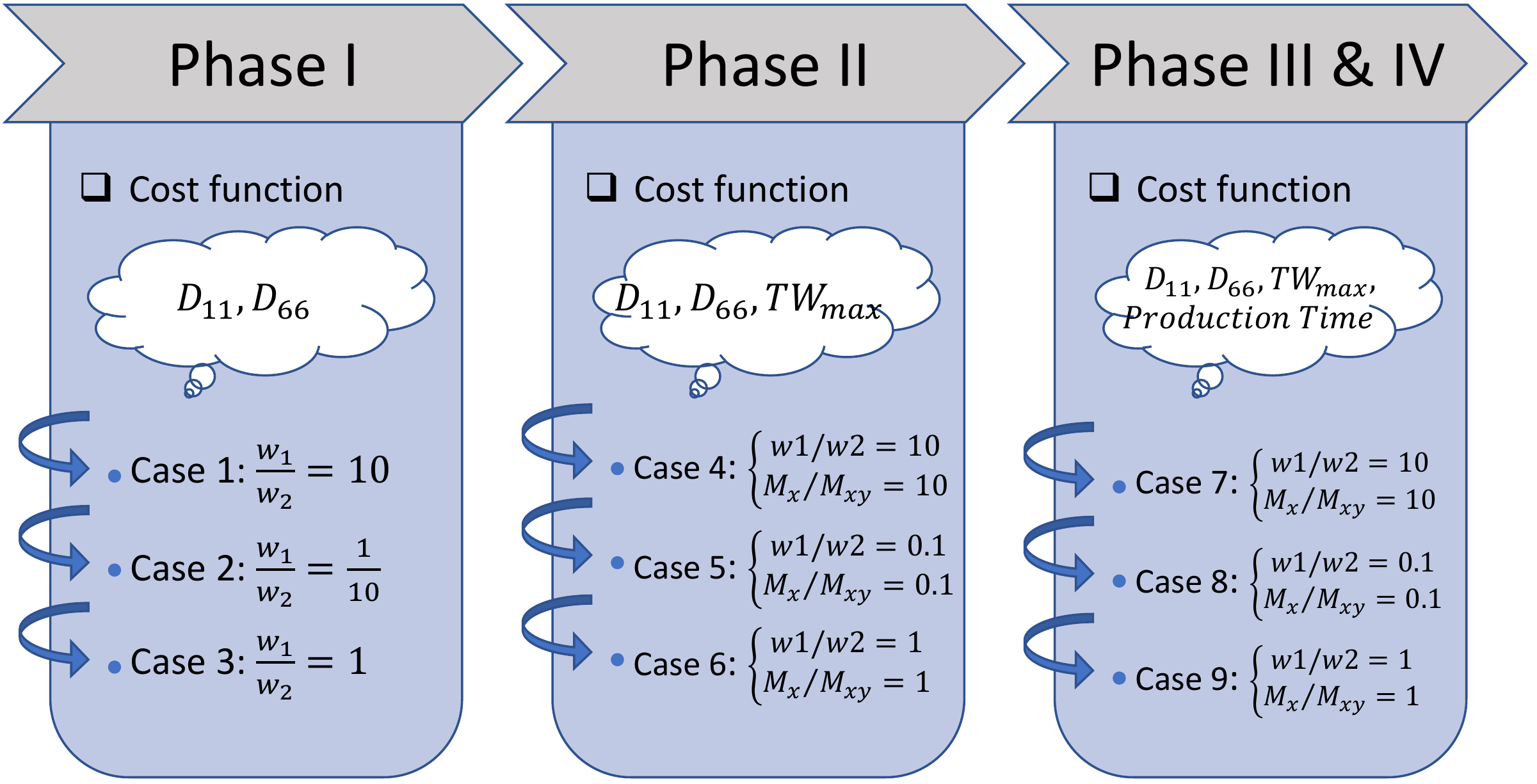}
\caption{Four phases of optimization}
\label{opt_cases}
\end{figure}

%(Comparison in 8-ply case and 32-ply case, in both cases, our results are better in terms of D11,D66,TW and manufacturing time)
% \jk{(We may want to insert a column in Table 2, grouping cases into different phases. And here, at the beginning of the section, we describe briefly these phases. E.g., In phase I, we will analyze stiffness only cases. In phase II, stiffness and XX, and in phase III, finally, all parameters including X, Y, and Z. Oh, in fact, I just found that you have this beautiful figure 5. You should mention that in the paragraph.)}
We break down the optimization study in 3 phases to isolate the influence of input parameters. In phase I, we only include the bending and torsional stiffness in the objective function. In phase II, we expand the objective function to contain the Tsai-Wu index to account for the influence of the initial ply failure. In phase III, we add the manufacturing time of a laminate in the objective function to include the manufacturing efficiency in the selection of the optimal laminate. Lastly, we extended the phase III optimization with 32 plies. This 4-phase optimization process is visually summarized in Fig. \ref{opt_cases}.
Phase I of the optimization focuses on the stiffness of the laminate (see Fig. \ref{opt_cases} for four phases of simulations conducted in this study). The results of the phase I optimization are summarized in Table \ref{tab:Table_result8ply_case123}. Case 1 in which the weight corresponding to $D_{11}$ is more substantial compared to the one related to $D_{22}$ has the optimal layup that is only consisted of $0^{\circ}$ ply. This result is convincing that the optimizer is on the right track considering the $0^{\circ}$ lamina provides the most bending resistance. Case 1 has the largest $D_{11}$ value in phase I. Case 2, on the other hand, produces the laminate with $\pm44^{\circ}$ plies as an optimal result. This is also a convincing result considering that the $\pm45^{\circ}$ plies provide the most resistance against a torsional load. Case 2 has the largest $D_{22}$ value in phase I. For case 3, in which both $D_{11}$ and $D_{66}$ weighed equally, the optimizer produced the laminate with $\pm34^{\circ}$ plies which can be seen as a compromise between case 1 and case 2 to achieve a balanced performance on both bending and torsional rigidity. The optimization result for phase I is consistent with the objective of each condition. We provide the $D_{11}$ and $D_{22}$ values of quasi-isotropic laminate for comparisons. It is theoretically apparent for case 1 and case 2 to have a better performance under bending and torsional load, respectively, compared to the quasi-isotropic case. We want to highlight the result of case 3. We see that by having the ply angle away from $0^{\circ}$, the optimal laminate loses the bending resistance in a significant amount compared to case 1. However, it is still much comparable to quasi-isotropic laminate. On the other hand, an increase in torsional resistance is superior compared to the one of quasi-isotropic laminate. Therefore, the overall gain in the structural performance of case 3 by compensating a decrease in $D_{11}$ is desirable over quasi-isotropic laminate for a structure going under torsional loading.

\begin{figure} [H]
\center
\includegraphics[scale=0.15]{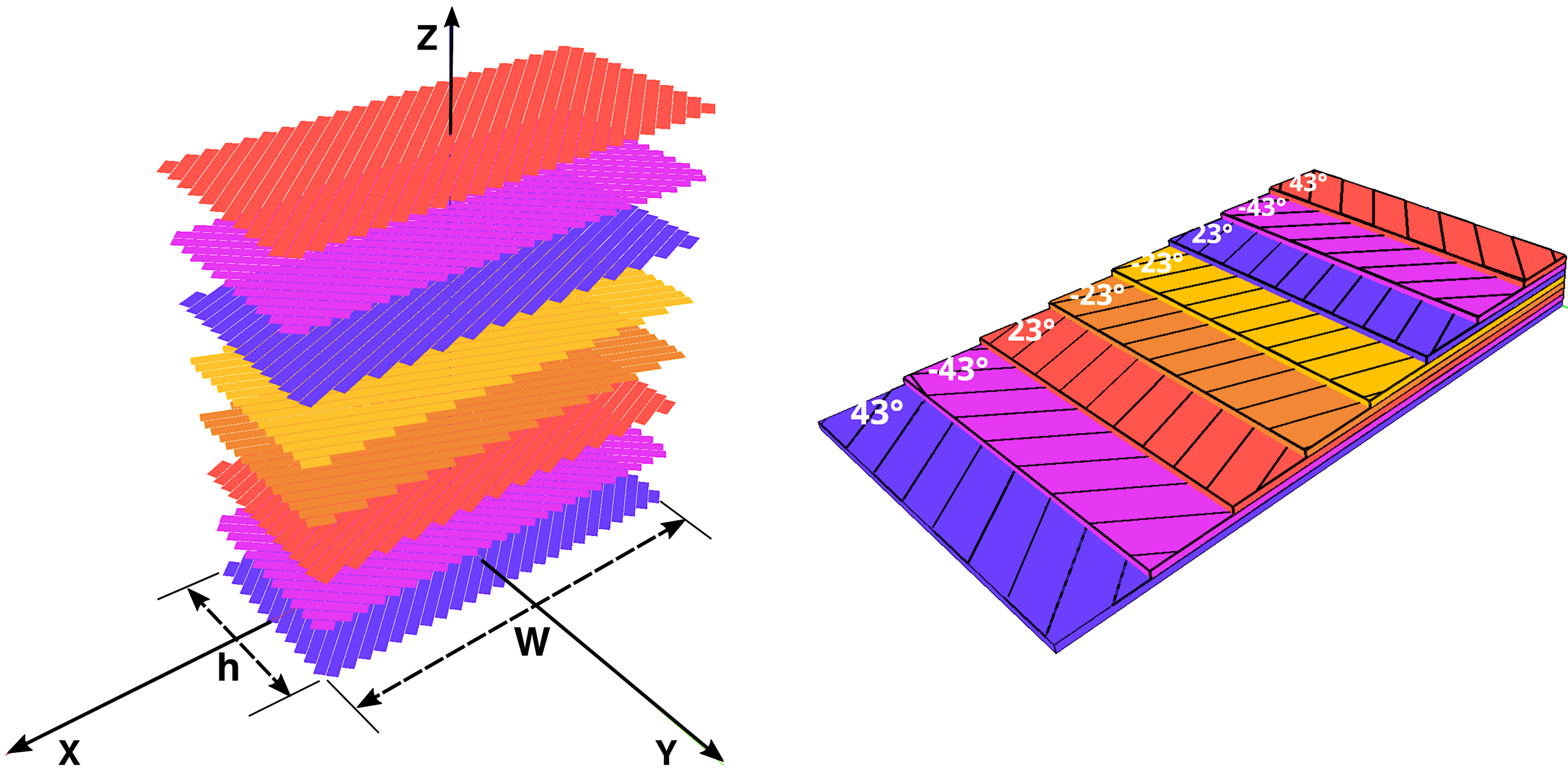}
\caption{Optimal laminate configuration for Case 6}
\label{GUI_case6}
\end{figure}

Phase II of the optimization additionally includes the evaluation of the initial ply failure. The result of the optimization is shown in Table \ref{tab:Table_result8ply_case456}. The maximum Tsai-Wu value for the quasi-isotropic laminate, $TW_{qimax}$ under the same loading condition is computed for each case. Lower $TW_{max}$ means that the evaluation point on the ply susceptible to an initial failure is farther away from the omni-envelop; therefore, a laminate with the lower ratio, $R_{TW} = TW_{max}/TW_{qimax}$ has better initial damage resistance. As can be noted, changes in ply angles for case 4 and case 5 are minimal from case 1 and case 2, respectively. On the contrary, ply angles of the optimal laminate for case 6 are noticeably different from those for case 3. In phase I, each optimal laminate has a single set of ply angle, $\pm\theta$, forming a cross-ply. However, case 6 have two distinct sets of ply angle: $\pm43^{\circ}$ (closer to $\pm45^{\circ}$ from $\pm32^{\circ}$) and $\pm23^{\circ}$ (closer to $0^{\circ}$ from $\pm32^{\circ}$). The layup visualization of optimal laminate for case 6 is shown in Fig. \ref{GUI_case6}. It appears that the optimizer used this variation in each set of ply angle as a mechanism to keep the balanced ratio of $D_{11}$ and $D_{66}$ and to achieve the low maximum Tsai-Wu value. Compared to the quasi-isotropic laminate, the optimal laminate for case 6 has improved $D_{66}$ and resistance to initial ply failure with further compensation on $D_{11}$.

In phase III, we lastly added manufacturing time in the cost function during the optimization process. From the development of tape-layering simulation, we noticed that shallower the angles are shorter the manufacturing times are for the chosen plate geometry. Therefore, optimizer tends to force the laminate to have a shallow angle as can be noted in Table \ref{tab:Table_result8ply_case789}. Ply angles of the optimal laminate for case 7 are $0^{\circ}$ as expected. Major changes are shown in both case 8 and 9. Each optimal case has a pair of $0^{\circ}$ to reduce the total layup time for each laminate. In case 8, the pair of $\pm45^{\circ}$ keeps the torsional resistance comparable to case 5, and the optimal laminate with double paired ply angles outperforms the quasi-isotropic laminate in every aspect including the manufacturing time except the bending resistance. However, considering the torsional dominant loading, this compensation is acceptable. In case 9, the resultant layup has $\pm 35^{\circ}$ pair, which are shallower compared to $\pm 43^{\circ}$ pair in case 6. There is a slight reduction in the torsional resistance, $D_{66}$; however, the increase in the bending resistance, $D_{11}$ is significant. Shallow angle pairs in the optimal laminate in case 9 decrease the damage resistance compared to the quasi-isotropic layup; however, the reduction in the manufacturing time is significant along with the well-balanced ratio between bending and torsional resistance under mixed loading condition. For this reason, it makes the layup composition highly desirable over a conventional layup for designing well-balanced composite structures. As can be noticed, the layup composition for each optimal case is unconventional and deviates from the typical quasi-isotropic layup. 8-ply optimization restricts the ply angle variety up to two plus-minus pair, $\pm \theta$, by enforcing symmetry and balance in the layup formulation. From the observation, we expect to increase in both mechanical and manufacturing performance for optimal laminates by increasing ply angles in the laminate.

We extend the optimization study in phase III to phase IV by increasing the number of layup composition angles to 32. Enforcing the symmetry and balanced layup provides 8 ply angle pairs, $\pm \theta$. The results of the optimization are shown in Table \ref{tab:Table_result32ply_case101112}. As can be noted, optimizer tends to have a wider selection of ply angle instead of a conventional set of choices ($0^{\circ}, 90^{\circ}, \pm 45^{\circ}$) to improve the balance in structural and manufacturing performance. General trends shown with 8-ply optimization stay the same for 32-ply optimization. Under the bending-dominant loading condition (case 10), the optimal laminate consists of $0^{\circ}$ angles. These results are convincing because $0^{\circ}$ lamina provides high bending resistance, least layup time, and high resistance in initial damage under the bending dominant loading condition. When the torsional loading is relatively more significant as in case 11 and 12, ply angles in layup configuration starts to diversify. Under the torsional-dominant loading condition (case 11), the optimal laminate has 6 pairs of plus-minus ply angles closely aligned in 45 degrees. These angle pairs provide superior torsional stiffness and initial failure resistance compared to the quasi-isotropic laminate. Furthermore, two $0^{\circ}$ pairs keep the manufacturing time almost the same as the quasi-isotropic laminate. Lastly, similar diversity in the ply angles is shown in the optimal laminate in case 12. Ply angles are shallower compared to case 11, and there are $30^{\circ}$ pairs. This difference leads to significant improvement in bending stiffness and manufacturing time. Although each optimal laminate exhibits better performance considering all the aspects compared to the quasi-isotropic laminate in their loading conditions, we particularly would like to highlight the optimal laminated obtained in case 12. It surprisingly outperforms the quasi-isotropic laminate in every aspect. Even the torsional resistance is comparable to the one in case 11. This makes the notion that the quasi-isotropic laminate provides the most well-balanced layup configuration under an arbitrary loading type questionable. Lastly, we would like to point out that the layup configurations before a symmetry plane for the case 11 and 12 resemble Bouligand-like structure, which has been extensible studied for bio-inspired materials and shown to be effective in providing outstanding mechanical behaviors such as high impact resistance and damage tolerance \cite{apichattrabrut2006helicoidal,Zav12,Zav14,shang2016crustacean, liu2018failure, mencattelli2019realising}.  

% check head speed or manufacturing time 
% Last case check D66

\section{Conclusions}
We developed a ply-drop module to simulate the manufacturing of a rectangular composite plate using the Automated Fiber Placement machine and the structural module to evaluate structural performance under the combination of bending and torsional loading. We utilized the Bayesian optimization method to find the optimal layup composition for three different loading conditions. In phase I, the optimization process only considers a bending resistance and a torsional resistance. In phase II, we included the maximum Tsai-Wu index in the cost function to examine how the initial ply failure estimate affects the optimization results in phase I. Lastly, we added the manufacturing time in the cost function to account for the manufacturability of the composite layup in the optimization process. In conclusion, We provide the following observations: 
\begin{enumerate}
    \item Two sets of balanced ply angles, $\pm \theta$ can provide outperforming mechanical behaviors including initial damage resistance under various loading conditions. This implies that the conventional quasi-isotropic laminate is not the optimal configuration. This finding supports the efficacy of the `double-double' laminate proposed by Tsai et al. \cite{Tsai2019}.
    \item When the manufacturing time is included in the optimization process, optimal laminates tend to possess shallower ply angles that reduce the layup time. However, this can reduce the mechanical performance in the optimal laminates compared to the ones that do not consider manufacturability. 
    \item For the thicker laminates, a wider range and higher variety in the ply angle selection are observed. Moreover, Bouligand-like ply angle progression before the symmetry plane is shown for some optimal laminates. 
    \item As a notable result, optimal laminate for the case of the 32-ply laminate outperforms the conventional quasi-isotropic laminate in both mechanical and manufacturing sides.  
    \item This study shows how Bayesian optimization works well with multimodal functions as a derivative-free method. The optimization scheme demonstrates the search of design space without having the information about the cost function a priori, thanks to the statistical surrogate model and the policy of exploration and exploitation. This versatile algorithm can further be applied to the design of advanced structures with an efficient manufacturing process.
\end{enumerate}

Though this study accounted for a simple set of parameters, such as stiffness, strength, and manufacturing time, a more comprehensive combination of design parameters can be considered. This includes unique failure modes of shallow angle plies such as splitting, and the susceptibility of laminates to manufacturing imperfections (i.e., automated fiber placement machine-prone defects, gaps, and overlaps). For future works, we can also regard many aspects of the manufacturing of composite materials into consideration. For example, the homogenization feature in nonsymmetric double-double layups \cite{Tsai2019} is important because it allows no requirement for mid-plane symmetry, single ply drop for weight saving, 1-axis layup, and so on. Also, future applications are expected to include the analysis of more complex geometries and situations such as buckling, vibration, and warpage with the finite element analysis and detailed simulation of AFP machines.
% \ky{Here I tried to address Dr. Tsai's comments}

% Koshiro wrote (transfeered to the enumarated poits):
% Bayesian optimization has a great potential to work as a class of derivative-free methods where we do not have access to the derivatives of the quantities of interest. Also, Bayesian optimization works well with multimodal functions. Both are the cases for the problems in the design of advanced material and structures, and manufacturing of them. Future applications of the Bayesian optimization framework to the design of composite materials with the production via AFP machines are expected to include the analysis of more complex geometries with the finite element analysis, and detailed simulation of AFP machines. Furthermore, considering the manufacturing imperfection to the optimization process is also important.

%% The Appendices part is started with the command \appendix;
%% appendix sections are then done as normal sections
%% \appendix

%% \section{}
%% \label{}

%% If you have bibdatabase file and want bibtex to generate the
%% bibitems, please use
%%
 % \bibliographystyle{elsarticle-num}
 % \bibliography{carbonbib}

%% else use the following coding to input the bibitems directly in the
%% TeX file.

\section*{Declaration of competing interest}
The authors declare that they have no known competing financial interests or personal relationships that could have appeared to influence the work reported in this paper. 

\section*{CRediT authorship contribution statement}
\textbf{Koshiro Yamaguchi:} Conceptualization, Methodology, Investigation, Software, Visualization, Writing- Original draft preparation. \textbf{Sean E. Phenisee:} Conceptualization, Methodology, Investigation, Software, Visualization, Writing- Original draft preparation. \textbf{Zhisong Chen:} Investigation, Software, Visualization. \textbf{Marco Salviato:} Conceptualization, Methodology, Supervision, Writing - Review \& Editing, Funding acquisition.  \textbf{Jinkyu Yang:} Conceptualization, Methodology, Supervision, Writing - Review \& Editing, Funding acquisition.

\section*{Acknowledgments}
We thank Professor Tsai at Stanford University for helpful discussion. We are grateful for the financial support by the Joint Center for Aerospace Technology Innovation (JCATI) grant. K.Y. is supported by the Funai Foundation for Information Technology. 

%\section*{References}
%\begin{thebibliography}{00}
%\linespread{0.5} \small
%\setlength{\parskip}{0pt}
%\setlength{\itemsep}{0pt plus 0.3ex}
% example reference
%\bibitem{VanMat15}
%F. Van Der Klift, Y. Koga, A. Todoroki, M. Ueda, Y. Hirano, R. matsuzaki, 3D printing of continuous carbon fibre reinforced thermo-plastic (CFRTP) tensile test specimens, J. Compos. Mater, 6, (2015), 18--22.
%\end{thebibliography}

\bibliographystyle{elsarticle-num}
\bibliography{main}
%\addbibresource{JAFPref.bib}

\newpage
\section*{List of figures}

%\begin{figure} [H]
%\center
%\includegraphics[scale=0.4]{Figures/manufacturing_model.pdf}
%\caption{Modeling the tape-layering paths for a rectangular plate geometry}
%\label{manufacturing_model}
%\end{figure}

%\begin{figure} [H]
%\center
%\includegraphics[scale=0.65]{Figures/schematics.pdf}
%\caption{Lamina coordinate system and laminate schematics}
%\label{coordinate_system}
%\end{figure}

% \begin{figure} [H]
% \center
% \includegraphics[scale=0.8]{Figures/optim_process_2d_8ply_case6.pdf}
% \caption{Iterations within the optimization}
% \label{figure_iteration}
% \end{figure}

% \begin{figure} [H]
% \center
% \includegraphics[scale=0.1]{Figures/GUI_quasi.png}
% \caption{GUI: Quasi-isotropic layup}
% \label{opt_cases}
% \end{figure}

\newpage
\section*{List of tables}

\begin{table}[H]
    \caption{\textbf{Properties of the AFP machine and composite material}}
    \center
    \begin{tabular}{|l|l|l|l|l|l|}
        \hline
        $W$ (width) & 7000 [mm] & $E_1$ & 138 [GPa] & $X_t$ & 3250 [MPa]\\ \hline
        $h$ (length)& 3000 [mm] & $E_2$ & 10 [GPa] & $X_c$ & 1600 [MPa]\\ \hline
        tape width & 10 [mm] & $\nu _{12}$ & 0.34 & $Y_t$ & 62 [MPa]\\ \hline
        head speed & 500 [mm/s] & $\nu_{21}$ & $\frac{E_2}{E_1}\nu_{12}$ & $Y_c$ & 98 [MPa]\\ \hline
        head turning time & 10 [s]&$G_{12}$ & 7 [GPa] & $S$ & 75 [MPa]\\ \hline
        number of dropping tapes & 10 & $t$ (thickness) & 0.2 [mm] & - & -\\ \hline
    \end{tabular}
    \label{tab:Table_mechanicalproperties}
\end{table}

% \begin{table}[H]
%     \caption{\textbf{Loading conditions}}
%     \center
%     \begin{tabular}{|l|l|l|l|}
%         \hline
%         & Condition 1 & Condition 2 & Condition 3\\ \hline
%         $N_x$    & 0 & 0 & 0 \\ \hline
%         $N_y$    & 0 & 0 & 0 \\ \hline
%         $N_{xy}$ & 0 & 0 & 0 \\ \hline
%         $M_x$    & 10 [Nmm]& 1 [Nmm] & 10 [Nmm] \\ \hline
%         $M_y$    & 0       & 0       & 0 \\ \hline
%         $M_{xy}$ & 1 [Nmm] & 10 [Nmm] & 10 [Nmm]\\ \hline
%     \end{tabular}
%     \label{tab:Table_loadingconditions}
% \end{table}

% \begin{table}[H]
%     \caption{\textbf{Loading conditions}}
%     \center
%     \begin{tabular}{|l|l|l|l|l|}
%         \hline
%         & Condition 1 & Condition 2 & Condition 3 & Condition 4\\ \hline
%         $N_x$    & 0 & 0 & 0 & 0 \\ \hline
%         $N_y$    & 0 & 0 & 0 & 0 \\ \hline
%         $N_{xy}$ & 0 & 0 & 0 & 7 [MPa] \\ \hline
%         $M_x$    & 10 [Nmm]& 1 [Nmm] & 10 [Nmm] & 5 [kNm] \\ \hline
%         $M_y$    & 0       & 0       & 0 & 0 \\ \hline
%         $M_{xy}$ & 1 [Nmm] & 10 [Nmm] & 10 [Nmm] & 0 \\ \hline
%     \end{tabular}
%     \label{tab:Table_loadingconditions}
% \end{table}

\begin{table}[H]
    \caption{\textbf{Weight values}}
    \center
    \begin{tabular}{|l|l|l|l|l|l|l|}
        \hline
        Phase number& Case number& $w_1$ & $w_2$ & $w_3$ & $w_4$ & Loading condition\\ \hline
         \multirow{3}{*}{Phase I} & Case 1 & 10/11 & 1/11  & 0     & 0  & Condition 1\\ \cline{2-7}
         & Case 2 & 1/11  & 10/11 & 0     & 0  & Condition 2\\ \cline{2-7}
         & Case 3 & 1/2   & 1/2   & 0     & 0  & Condition 3\\ \hline
         \multirow{3}{*}{Phase II} & Case 4 & 5/11  & 1/22  & 1/2   & 0  & Condition 1\\ \cline{2-7}
         & Case 5 & 1/22  & 5/11  & 1/2   & 0  & Condition 2\\ \cline{2-7}
         & Case 6 & 1/4   & 1/4   & 1/2   & 0  & Condition 3\\ \hline
         \multirow{3}{*}{Phase III} & Case 7 & 20/55   & 2/55   & 3/10   & 3/10  & Condition 1\\ \cline{2-7}
         & Case 8 & 2/55   & 20/55   & 3/10   & 3/10 & Condition 2 \\ \cline{2-7}
         & Case 9 & 3/10  & 3/10 & 1/4   & 3/20 & Condition 3 \\ \hline
         \multirow{3}{*}{Phase IV} & Case 10 (32-ply) & 20/55   & 2/55   & 3/10   & 3/10  & Condition 1\\ \cline{2-7}
         & Case 11 (32-ply) & 2/55   & 20/55   & 3/10   & 3/10 & Condition 2 \\ \cline{2-7}
        % case12 (32-ply) & 1/5   & 1/5  & 3/10   & 3/10 & Condition 3 \\ \hline
         & Case 12 (32-ply) & 3/10  & 3/10 & 1/4   & 3/20 & Condition 3 \\ \hline
        % case14 (32-ply) & 1/5   & 3/10 & 2/5    & 1/5 & Condition 3 \\ \hline
        % case5 & 0.49  & 0.01  & 0.05  & 0.45 \\ \hline
    \end{tabular}
    \label{tab:Table_weights}
\end{table}

\begin{table}[H]
    \caption{\textbf{Comparison of the optimized result(8-ply) and quasi-isotropic laminate in phase I (case1, case2 and case3)}}
    \center
    \begin{tabular}{|c|c|c|c|c|}
        \hline
                     & Case 1 & Case 2 & Case 3 & Quasi-isotropic\\ \hline
        Layup angles & $[0,0,0,0]_s$ & $[44,-44,44,-44]_s$ & $[34,-34,34,-34]_s$ & $[0, 90, 45, -45]_s$\\ \hline
        $D_{11}$     & 47.5 [Nm] & 16.5 [Nm] & 25.3 [Nm]& 30.4 [Nm] \\ \hline
        $D_{66}$     & 2.39 [Nm] & 12.1 [Nm] & 10.8 [Nm]& 3.61 [Nm]  \\ \hline
    \end{tabular}
    \label{tab:Table_result8ply_case123}
\end{table}

\begin{table}[H]
    \caption{\textbf{Comparison of the optimized result(8-ply) and quasi-isotropic laminate in phase II (case4, case5 and case6)}}
    \center
    \begin{tabular}{|c|c|c|c|c|}
        \hline
                     & Case 4 & Case 5 & Case 6 & Quasi-isotropic\\ \hline
        Layup angles & $[0,0,1,-1]_s$ & $[46,-46,44,-44]_s$ & $[43,-43,23,-23]_s$ & $[0, 90, 45, -45]_s$\\ \hline
        $D_{11}$     & 47.5 [Nm] & 15.1 [Nm] & 19.6 [Nm] & 30.4 [Nm] \\ \hline
        $D_{66}$     & 2.39 [Nm] & 12.1 [Nm] & 11.5 [Nm] & 3.61 [Nm] \\ \hline
        $R_{TW}$   & 68.4\%   & 77.6\% & 81.9\% & - \\ \hline
    \end{tabular}
    \label{tab:Table_result8ply_case456}
\end{table}

\begin{table}[H]
    \caption{\textbf{Comparison of the optimized result(8-ply) and quasi-isotropic laminate in phase III (case7, case8 and case9)}}
    \center
    \begin{tabular}{|c|c|c|c|c|}
        \hline
                     & Case 7 & Case 8 & Case 9 & Quasi-isotropic\\ \hline
        Layup angles & $[0,0,0,0]_s$  & $[45,-45,0,0]_s$ & $[35,-35,0,0]_s$ & $[0, 90, 45, -45]_s$\\ \hline
        $D_{11}$     & 47.5 [Nm] & 19.7 [Nm] & 27.3 [Nm] & 30.4 [Nm] \\ \hline
        $D_{66}$     & 2.39 [Nm] & 10.9 [Nm] & 9.93 [Nm] & 3.61 [Nm] \\ \hline
        $R_{TW}$   & 68.6\%   & 82.5\% & 104\% & - \\ \hline
%        $Time$ & 5760 [s] & 7456 [s] & 7220 [s] & 8256 [s]\\ \hline
        $Time$ & 96 [min] & 124 [min] & 120 [min] & 138 [min]\\ \hline
    \end{tabular}
    \label{tab:Table_result8ply_case789}
\end{table}

\begin{table}[H]
    \caption{\textbf{Comparison of the optimized result(32-ply) and quasi-isotropic laminate (case10, case11 and case12)}}
    \center
    \begin{tabular}{|c|c|c|c|c|}
        \hline
      & Case 10 & Case 11 & Case 12 & Quasi-isotropic \\ \hline
\begin{tabular}[c]{@{}c@{}}Layup\\ angles\end{tabular} & $0_{32}$       & \begin{tabular}[c]{@{}c@{}}[$\pm$ 50, $\pm$45, $\pm$45, $\pm$43, \\ $\pm$43, $\pm$38, $\pm$0, $\pm$0]$_{s}$ \end{tabular} & \begin{tabular}[c]{@{}c@{}}[$\pm$43, $\pm$40, $\pm$34, $\pm$32, \\ $\pm$31, $\pm$0, $\pm$0, $\pm$0]$_{s}$\end{tabular} & \begin{tabular}[c]{@{}c@{}}[$\pm$0, $\pm$90, \\ $\pm$45, $\pm$45]$_{s}$\end{tabular} \\ \hline
        $D_{11}$    &  3.04 [kNm] & 1.00 [kNm] & 1.47 [kNm] & 1.36 [kNm] \\ \hline
        $D_{66}$    &  152 [Nm] & 760 [Nm] & 699 [Nm] & 451 [Nm]\\ \hline
        $R_{TW}$  &  29.8$\%$   & 77.6$\%$ & 75.3$\%$ & - \\ \hline
%        $time$ &  $2.30\times 10^4$ [s] & $3.31 \times 10^4$ [s] & $3.05 \times 10^4$ [s] & $3.30 \times 10^4$\\ \hline
        $Time$ & 383 [min] & 551 [min] & 508 [min] & 550 [min]\\ \hline        
    \end{tabular}
    \label{tab:Table_result32ply_case101112}
\end{table}

\end{document}